\documentclass[lettersize,journal]{IEEEtran}

\IEEEoverridecommandlockouts
\usepackage{booktabs}
\usepackage{siunitx}
\usepackage{amsfonts}
\usepackage[dvips]{graphicx}
\usepackage{times}
\usepackage{cite}
\usepackage{amsmath}
\usepackage{cases}
\usepackage{array}
\usepackage{dsfont}
\usepackage{amssymb}

\usepackage{makecell}
\usepackage{stfloats}
\usepackage{booktabs}
\usepackage{graphicx}
\usepackage{subfigure}
\usepackage{pdfpages}
\usepackage{footnote}
\usepackage{soul}
\usepackage{array}
\usepackage{multirow}
\usepackage{bm}
\usepackage{empheq}
\usepackage{amsthm}
\usepackage{algorithm}
\usepackage{algorithmicx}
\usepackage{algpseudocode}
\usepackage{color}
\usepackage{diagbox}
\usepackage{caption}
\theoremstyle{plain}

\theoremstyle{definition}

\usepackage[colorlinks,bookmarksopen,bookmarksnumbered,citecolor=blue, linkcolor=blue, urlcolor=blue]{hyperref}

\let\oldReturn\Return
\renewcommand{\Return}{\State\oldReturn}

\usepackage{textcomp}
\usepackage{xcolor}
\begin{document}
	\title{Flow Matching-Based Semantic Communication With Time-Aware Structural Residual Injection}
	\vspace{-4pt}
	
	\vspace{-25pt}\author{\IEEEauthorblockN{$\text{Chengyang Liang}$, $\text{Dong Li}, ~\IEEEmembership{Senior Member,~IEEE}$}
		\vspace{-22pt}
		\thanks{Chengyang Liang and Dong Li are with the School of Computer Science and Engineering, Macau University of Science and Technology, Macau, China (e-mail: 3240006992@student.must.edu.mo; dli@must.edu.mo).}}
	\vspace{-20pt}

\maketitle

\begin{abstract}
	Generative semantic communication (SemCom) enables low-bandwidth image transmission by reconstructing images from semantic descriptions. However, text-only schemes often suffer from poor structural fidelity due to the stochastic nature of generative models, while still requiring high transmission payload and inference latency. This letter proposes a dual-information-flow SemCom framework based on Flow Matching. At the transmitter, the most semantically matched prompt is selected from multiple candidate descriptions, while an ultra-low-dimensional structural residual velocity vector is extracted from the inversion trajectory of the original image. At the receiver, a Time-Aware Recursive Residual Injection (TARRI) network progressively releases structural corrections along the denoising trajectory using the transmitted residual vector as structural memory. The proposed framework transmits only prompt embedding and a compact residual vector, achieving robust high-fidelity image reconstruction under extremely low payload. Experimental results demonstrate that the proposed framework significantly improves both structural fidelity and perceptual quality over existing text-driven semantic communication schemes.

\end{abstract}


\IEEEpeerreviewmaketitle

\section{Introduction}
\label{sec:Introduction}

\IEEEPARstart{T}{he} rapid proliferation of Internet of Things (IoT) visual devices has led to explosive growth in wireless image traffic. However, transmitting high-fidelity visual information over bandwidth-constrained wireless links remains highly challenging for resource-limited IoT systems. Conventional image transmission frameworks mainly rely on pixel-level compression, whose performance degrades significantly under extremely low bit-rate conditions. Recently, generative semantic communication (SemCom) has emerged as a promising paradigm for low-bandwidth image transmission \cite{GenAIsemcomLiang}, \cite{11569078}. Instead of transmitting compressed pixels, the transmitter conveys semantic descriptions, while the receiver reconstructs images using generative models. This paradigm reduces transmission overhead by shifting from pixel fidelity to semantic consistency. Recent generative SemCom frameworks can be broadly categorized into two streams: latent transmission schemes \cite{DeepJSCC} that directly transmit compressed latent representations from generative models, and diffusion inversion-based methods \cite{Diffsc} that leverage the reverse trajectory of diffusion models to reconstruct images from transmitted semantic information. More recently, prompt-guided approaches \cite{promptsc}, \cite{PromptguidedSC}, \cite{ImageGeneration} have been proposed to improve semantic consistency by transmitting textual descriptions alongside generative conditions. Nevertheless, existing text-driven SemCom frameworks suffer from a fundamental limitation: text prompts alone cannot sufficiently constrain structural details in the generated image. Due to the stochastic nature of generative models, different realizations under the same prompt may exhibit large variations in object layout, contours, and textures, resulting in poor structural fidelity \cite{TaskVQA}, \cite{ImageGeneration}, \cite{LaMoSC}.

To compensate for the structural ambiguity caused by the stochastic nature of generative models, existing methods usually transmit additional structural information together with text prompts to better constrain the reconstruction process \cite{promptsc}. However, most transmitted auxiliary information remains essentially compressed pixel-domain or latent-domain representations, which still contain considerable redundancy and become fragile under extremely low bit-rate conditions. Moreover, directly reconstructing structural information from compact latent vectors neglects the inherently time-varying structural demands across the generative trajectory \cite{PromptguidedSC}. To the best of our knowledge, Flow Matching has not yet been explored for generative SemCom, despite its appealing properties. Unlike diffusion-based models requiring hundreds of sampling steps, Flow Matching \cite{lipman2022flow} adopts a continuous-time ordinary differential equation (ODE) formulation with shorter integration paths for low-latency IoT image transmission. This motivates our investigation of a Flow Matching-based SemCom framework.

Motivated by the low structural fidelity, high transmission payload, and long inference time of existing generative SemCom methods, and more fundamentally, by the challenge of identifying which generation dynamics remain irrecoverable from semantic prompts and how such information should be allocated across different stages of the generative trajectory, this letter proposes a dual-information-flow SemCom framework based on Flow Matching. Unlike conventional latent transmission methods, the proposed framework transmits only prompt embeddings and the residual generation dynamics that cannot be explained by text semantics. Specifically, the generation process in Flow Matching is formulated as ODE integration within a velocity field, whose residual component naturally characterizes structural deviations between text-conditioned generation and the target image trajectory. Based on this property, the transmitter extracts an ultra-low-dimensional structural residual velocity vector from the inversion trajectory of the original image, while the receiver progressively unfolds the transmitted structural information along the denoising trajectory. To efficiently utilize the transmitted residual information, we further propose a Time-Aware Recursive Residual Injection (TARRI) network. TARRI treats the residual vector as structural memory and recursively releases corrections across denoising steps for stage-aware refinement.

The main contributions of this letter are summarized as follows:
\begin{itemize}
	\item We propose a dual-information-flow generative SemCom framework that jointly transmits semantic prompt embeddings and compact structural residual dynamics for high-fidelity image reconstruction under extremely low payload.
	
	\item We propose a structural residual velocity extraction mechanism based on Flow Matching inversion, which identifies and encodes the generation dynamics unrecoverable from semantic prompts into a low-dimensional representation.
	
	\item We propose a TARRI network that progressively allocates structural residual dynamics across different denoising stages through recursive temporal decoding, significantly improving structural fidelity and robustness under noisy wireless channels.
	
	\item Experimental results demonstrate that the proposed framework outperforms representative SemCom baselines in terms of reconstruction quality with substantially lower transmission payload and inference latency.
\end{itemize}

\vspace{-12pt}
\section{System Model}
\label{sec:SystemModel}
\vspace{-4pt}

\begin{figure}[tp]
	\vspace{-10pt}
	\centering
	\includegraphics[scale=0.6]{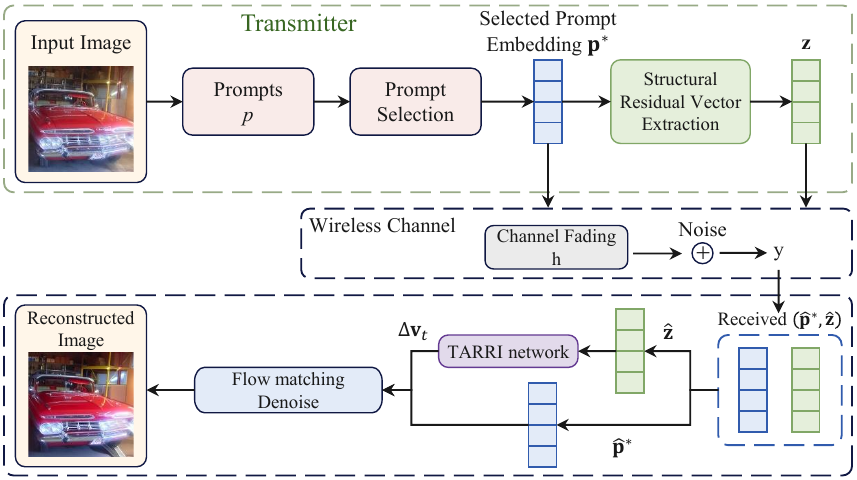}
	\caption{Overview of the proposed semantic communication framework.}
	\label{fig:system}
	\vspace{-16pt}
\end{figure}

Consider an image SemCom system over a wireless channel. Let $\mathbf{I}\in\mathbb{R}^{H\times W\times 3}$ denote the source image. The transmitter first generates multiple candidate semantic descriptions $\mathbf{p}$ and selects the optimal prompt embedding $\mathbf{p}^{*}$ through semantic matching. Meanwhile, the image is projected into the Flow Matching latent trajectory space to extract compact structural residual dynamics.

The transmitter encoder can be abstractly formulated as

\vspace{-4pt}
\begin{equation}
	\vspace{-4pt}
(\mathbf{p}^{*},\mathbf{z}) = f_{\mathrm{enc}}(\mathbf{I};\alpha),
\end{equation}

\noindent where $\mathbf{p}^{*}$ denotes the selected semantic prompt embedding, $\mathbf{z}\in\mathbb{R}^{d}$ denotes the low-dimensional structural residual vector, and $\alpha$ represents the encoder parameters.

The encoded features are further modulated into complex-valued transmit symbols and delivered over a Rayleigh fading wireless channel impaired by additive white Gaussian noise (AWGN). The channel model is expressed as

\vspace{-6pt}
\begin{equation}
	\vspace{-6pt}
	\mathbf{y}=h\mathbf{x}+\mathbf{n}
\end{equation}

\noindent where $\mathbf{x}\in\mathbb{C}^{N}$ and $\mathbf{y}\in\mathbb{C}^N$ denote the transmitted and received semantic information, $h\sim\mathcal{CN}(0,1)$ represents the Rayleigh fading channel coefficient, and $\mathbf{n}\sim\mathcal{CN}(0,\sigma_{n}^{2}\mathbf{I})$ denotes the complex AWGN vector. In practice, perfect channel state information (CSI) may be unavailable at the receiver. To account for channel estimation errors, we model the estimated channel coefficient as $\hat{h}=h+\epsilon$, where $\epsilon\sim\mathcal{CN}(0,\sigma_e^2)$ denotes the estimation error with variance $\sigma_{e}^2$. The quantized features are binary phase shift keying (BPSK)-modulated into $N$ symbols for transmission, with $N$ determined by the total payload length.

At the receiver, the semantic decoder reconstructs the image conditioned on both the received prompt embedding $\hat{\mathbf{p}}^*$ and the vector $\hat{\mathbf{z}}$. The reconstruction process is formulated as

\vspace{-6pt}
\begin{equation}
	\vspace{-6pt}
\hat{\mathbf{I}}
=
f_{\mathrm{dec}}
(\hat{\mathbf{p}}^*,\hat{\mathbf{z}};\rho),
\end{equation}

\noindent where $\rho$ denotes the decoder parameters.

Unlike conventional latent transmission schemes, the proposed method communicates only prompt embeddings and structural residual dynamics unrecoverable from semantic text conditions. The transmitted residual vector is progressively injected into the generative trajectory during denoising, enabling high-fidelity reconstruction under low transmission overhead.

\vspace{-7pt}
\section{Proposed Semantic Communication Framework}
\label{sec:architecture}

The proposed framework jointly transmits semantic content and structural residual dynamics under extremely low payload. As illustrated in Fig.~\ref{fig:system}, the transmitter first selects the semantic prompt with the highest semantic consistency to preserve semantic-level information during transmission. Based on the selected prompt, compact structural residual dynamics are further extracted from the Flow Matching inversion trajectory to compensate for structural details unrecoverable from text semantics alone. At the receiver, the transmitted residual information is progressively unfolded along the denoising trajectory through the proposed Time-Aware Recursive Residual Injection (TARRI) network.

\vspace{-8pt}
\subsection{Optimal Prompt Selection}
\vspace{-2pt}

$\mathbf{I}$ denotes the source RGB image. A pretrained Bootstrapping Language-Image Pre-training (BLIP) image-captioning model first generates $K$ candidate textual descriptions from the source image, which are then encoded into candidate prompt embeddings $\{p_i\}_{i=1}^{K}$ via the Contrastive Language-Image Pre-training (CLIP) text encoder. To improve semantic consistency between the transmitted prompt embedding and the source image, the optimal prompt embedding is selected based on cosine similarity

\vspace{-6pt}
\begin{equation}
	\vspace{-6pt}
\mathbf{p}^{*}
=
\arg\max_{\mathbf{p}_i}
\frac{
	\phi_I(\mathbf{I})^\top
	\phi_T(p_i)
}{
	\|\phi_I(\mathbf{I})\|_2
	\|\phi_T(p_i)\|_2
},
\end{equation}

\noindent where $\phi_I(\cdot)$ and $\phi_T(\cdot)$ denote the CLIP image and text encoders, respectively. The selected prompt embedding $\mathbf{p}^{*}$ serves as the semantic condition for subsequent image generation.

\vspace{-10pt}
\subsection{Structural Residual Dynamics Extraction}
\vspace{-4pt}

Semantic prompts preserve high-level information but are insufficient to fully constrain structural details during stochastic generation. The Flow Matching inversion process begins with a source image $\mathbf{I}\in\mathbb{R}^{H\times W\times 3}$, which is first encoded into the latent space via a pretrained variational autoencoder (VAE) encoder $\mathbf{l}_0 = \mathcal{E}_{\mathrm{VAE}}(\mathbf{I})$, where $\mathcal{E}_{\mathrm{VAE}}(\cdot)$ denotes the pretrained VAE encoder. The latent code is then progressively inverted along the full Flow Matching trajectory

\vspace{-8pt}
\begin{equation}
	\vspace{-8pt}
	\mathbf{l}_{t+\Delta t} = \mathbf{l}_t + \Delta t \, v_\theta(\mathbf{l}_t, t, \mathbf{p}^*),
\end{equation}

\noindent where $v_\theta(\cdot)$ denotes the pretrained Flow Matching velocity predictor, and $t$ indexes the inversion steps from the terminal latent $\mathbf{l}_T$ back to $\mathbf{l}_0$. Formally, the velocity field $v_\theta(\mathbf l_t,t,\mathbf p^*)$ is conditioned on the selected semantic prompt and therefore characterizes the prompt-guided generation trajectory. The structural residual is defined as the complementary trajectory information required to reproduce the source image beyond what can be recovered from this prompt-conditioned generation process.

In our framework, instead of transmitting or storing the full inversion trajectory, we extract only a compact structural residual vector from selected velocity states along the trajectory. Specifically, multiple sampled velocity states from the prompt-conditioned inversion trajectory are temporally mean-pooled and encoded by a lightweight two-layer multilayer perceptron (MLP) residual extractor $f_V(\cdot)$, yielding a compact 32-dimensional structural residual vector. In our implementation, $f_V$ takes 8 uniformly sampled velocity frames of shape $(8, 16, 64, 64)$ as input and is trained end-to-end with the image reconstruction loss, ensuring that the extracted residual captures prompt-irrecoverable structural information.

\vspace{-6pt}
\begin{equation}
	\vspace{-6pt}
	\mathbf{z} = f_\mathrm{V}(\{v_\theta(\mathbf{l}_t, t, \mathbf{p}^*)\}_{t=1}^{T}),
\end{equation}

\noindent where $\mathbf{z}\in\mathbb{R}^d$ denotes the compact structural residual representation, and $f_V(\cdot)$ aggregates the sampled velocity states into a low-dimensional residual vector.

\vspace{-10pt}
\subsection{Time-Aware Recursive Residual Injection}

\begin{figure}[htp]
	\vspace{-10pt}
	\centering
	\includegraphics[scale=0.65]{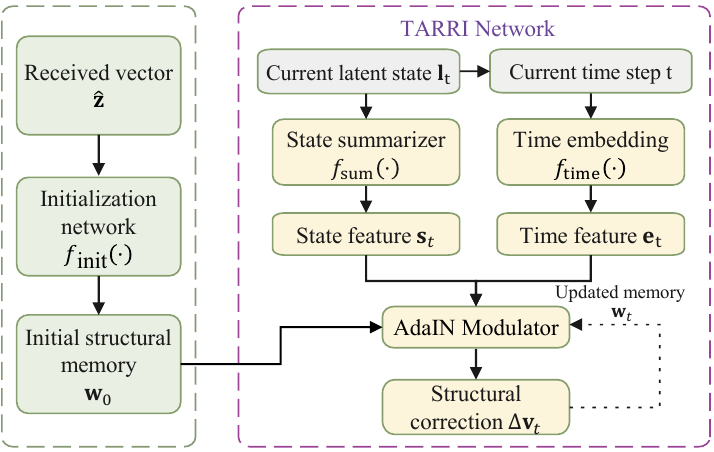}
	\caption{Architecture of the proposed TARRI network.}
	\label{fig:TARRI}
	\vspace{-10pt}
\end{figure}

At the receiver, directly recovering all structural details from a single residual vector is challenging due to the hierarchical nature of the denoising process. To address this issue, we propose a TARRI network that progressively unfolds the transmitted structural information along the reverse generation trajectory via lightweight temporal modulation, as illustrated in Fig.~\ref{fig:TARRI}. The network consists of three adaptive instance normalization (AdaIN) residual blocks with channel dimensions $\{16, 64, 32, 16\}$, taking the received residual vector, current latent state, and timestep embedding as conditioning inputs.

The structural residual vector $\mathbf{z}$ is transmitted only once and the received vector $\hat{\mathbf{z}}$ is first transformed into an initial structural memory $\mathbf{w}_0=f_{\mathrm{init}}(\hat{\mathbf{z}})$, where $f_{\mathrm{init}}(\cdot)$ denotes the initialization network and $\mathbf{w}_0$ represents the structural memory maintained throughout the denoising process.

During reverse generation, the current latent state $\mathbf{l}_t$ and timestep $t$ are encoded as $\mathbf{s}_t=f_{\mathrm{sum}}(\mathbf{l}_t)$ and
$\mathbf{e}_t=f_{\mathrm{time}}(t)$, where
$f_{\mathrm{sum}}(\cdot)$ and $f_{\mathrm{time}}(\cdot)$ denote the latent-state summarizer and temporal embedding network.

The state representation and temporal embedding are concatenated and recursively processed together with the previous structural memory

\vspace{-3pt}
\begin{equation}
	\vspace{-3pt}
	\mathbf{w}_t,\Delta\mathbf{v}_t=AdaIN([\mathbf{s}_t,\mathbf{e}_t],\mathbf{w}_{t-1}),
\end{equation}

\noindent where $\mathbf{w}_{t}$ denotes the updated structural memory and $\Delta\mathbf{v}_t$ denotes the structural velocity correction at timestep $t$.

The decoded correction is injected into the semantic velocity field predicted by the pretrained Flow Matching model

\vspace{-6pt}
\begin{equation}
	\vspace{-6pt}
	\tilde{\mathbf{v}}_t=v_\theta(\mathbf{l}_t,t,\hat{\mathbf{p}}^*)+\lambda\tanh(\gamma)\Delta\mathbf{v}_t,
\end{equation}

\noindent where $\lambda$ denotes the training-stage modulation factor and $\gamma$ is a learnable scaling parameter.

The latent state is then updated through reverse ODE integration

\vspace{-6pt}
\begin{equation}
	\vspace{-6pt}
	\mathbf{l}_{t-\Delta t}=\mathbf{l}_t-\Delta t\tilde{\mathbf{v}}_t.
\end{equation}

After completing the reverse integration process, the final latent representation is decoded through the pretrained VAE decoder $\hat{\mathbf{I}}=\mathcal{D}_{\mathrm{VAE}}(\hat{\mathbf{l}}_0)$, where $\mathcal{D}_{\mathrm{VAE}}$ denotes the pretrained VAE decoder, yielding the reconstructed image.

\vspace{-8pt}
\section{Numerical Experiments}
\label{sec:Experiments}
\vspace{-4pt}

\subsection{Experimental Settings}

Experiments are conducted on the ImageNet-512 dataset. We use a subset of ImageNet-512 consisting of 1,300 training images and 100 test images. All reported results are averaged over 50 independent channel trials per test image at each signal-to-noise ratio (SNR) point. The proposed framework employs the Flow Matching model as the generative backbone and CLIP for semantic similarity evaluation and prompt selection. The number of candidate prompt descriptions is set to $K=5$.The total transmission payload consists of the quantized prompt embedding plus the quantized residual vector. No additional channel coding is applied. The quantized bits are directly mapped to BPSK symbols for transmission over the physical channel. The system directly transmits the quantized prompt embedding rather than the raw text prompt or token sequence, since the embedding is the actual conditioning input to the generative model at the receiver. For structural residual extraction, the source image is first encoded into the latent space using the pretrained VAE encoder. The Flow Matching inversion process adopts 15 ODE steps, while the image reconstruction process employs 30 reverse denoising steps. During training, multiple velocity states sampled from the inversion trajectory are aggregated to construct the transmitted structural residual representation. For reference, the prompt embedding $\mathbf{p}^*$ is $768$-dimensional, the residual vector $\mathbf{z}$ has dimension $32$, the latent state $\mathbf{l}_t$ is in $\mathbb{R}^{16\times64\times64}$, the velocity fields $\mathbf{v}_t$ and $\Delta\mathbf{v}_t$ share the same shape $\mathbb{R}^{16\times64\times64}$, and the number of candidate prompts is $K=5$.

To evaluate transmission robustness under practical wireless environments, both AWGN and Rayleigh fading channels are considered. The SNR varies from -10 dB to 10 dB. The reconstruction performance is evaluated using three widely adopted metrics, including Peak SNR (PSNR), Structural Similarity Index Measure (SSIM), and CLIP Score. PSNR and SSIM evaluate pixel-level reconstruction fidelity, while CLIP Score quantifies semantic consistency between the reconstructed image and the original semantic content. All experiments are conducted on a workstation equipped with four NVIDIA V100 GPUs and an AMD EPYC 7742 CPU.

\vspace{-10pt}
\subsection{Performance Analysis}

\vspace{2pt}
\begin{figure}[tp]
	\vspace{-2pt}
	\centering
	\includegraphics[scale=0.13]{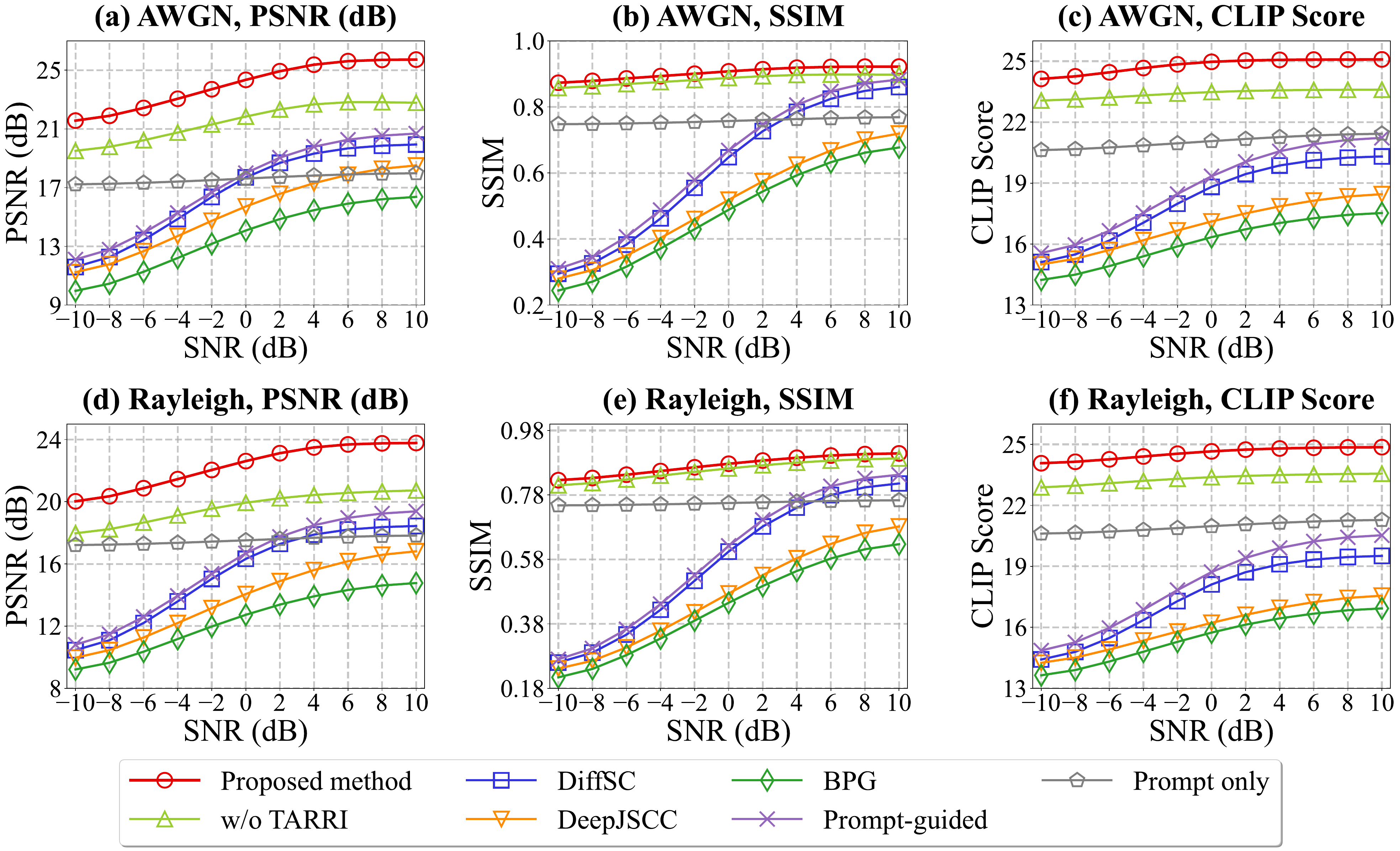}
	\caption{Performance comparison on the ImageNet-512 dataset under AWGN and Rayleigh fading channels.}
	\label{fig:comparison}
	\vspace{-3pt}
\end{figure}

\vspace{-3pt}
\begin{figure}[t]
	\centering
	\includegraphics[width=0.9\columnwidth]{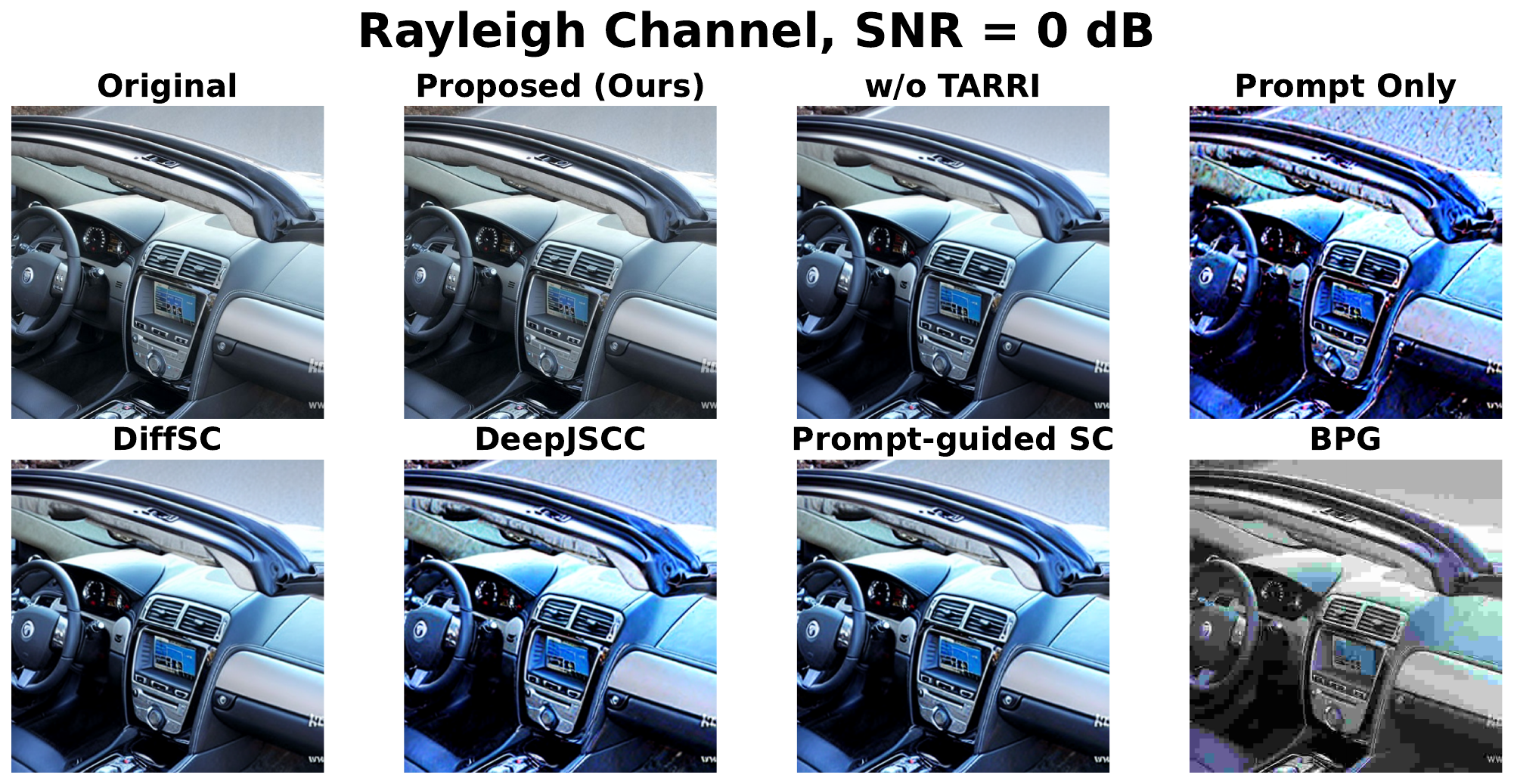}
	\caption{Visual comparison between different baselines under Rayleigh fading.}
	\label{fig:visual_comparison}
	\vspace{-12pt}
\end{figure}

To evaluate the effectiveness of the proposed method, we compare it with several representative baselines: DeepJSCC \cite{DeepJSCC}, DiffSC \cite{Diffsc}, Better Portable Graphics (BPG) and Prompt-guided SemCom method \cite{promptsc}. DeepJSCC and DiffSC represent conventional and diffusion-based SemCom approaches. BPG is a traditional compression baseline, and Prompt-guided \cite{promptsc} is a text-driven SemCom scheme.

Two ablation variants are considered. The \emph{Prompt Only} transmits only the prompt embedding without structural residual information. The \emph{w/o TARRI} retains residual transmission but replaces TARRI with an MLP that applies a fixed velocity correction across all denoising steps, removing the time-aware memory propagation.

Experimental results under both AWGN and Rayleigh fading channels are presented in Fig.~\ref{fig:comparison}. The proposed method consistently achieves the best performance across all metrics and channel conditions. Compared with other baselines, the proposed framework obtains higher PSNR, SSIM, and CLIP Scores, demonstrating its ability to simultaneously preserve structural fidelity and semantic consistency.

The ablation results further verify the effectiveness of the proposed design. Compared with the \emph{Prompt Only} scheme, introducing the structural residual vector significantly improves reconstruction quality, indicating that semantic prompts alone cannot adequately recover fine-grained structural details. Moreover, the \emph{w/o TARRI} variant consistently underperforms the proposed method, demonstrating that directly applying a fixed structural correction is less effective than the proposed recursive temporal decoding mechanism. By progressively releasing structural information along the denoising trajectory, TARRI enables more accurate structural reconstruction and perceptual refinement. The visual comparison in Fig.~\ref{fig:visual_comparison} corroborates the quantitative results, demonstrating that the proposed method effectively maintains structural fidelity and perceptual quality even under severe channel degradation.

\vspace{-4pt}
\begin{table}[t]
	\centering
	\caption{COMPLEXITY AND ABLATION STUDY}
	\label{tab:complexity_ablation}
	\resizebox{\columnwidth}{!}{%
		\begin{tabular}{lcccccc}
			\hline
			Method & Parameters & Payload & Inference Time & $d$ & $\lambda$ & PSNR@5dB \\
			& (M) & (bits) & (ms) & & & (AWGN) \\
			\hline
			Proposed (d=32, $\lambda$=1) & 4.48 & 6,400 & 68.2 & 32 & 1.0 & 25.68 \\
			d=16, $\lambda$=1 & 4.10 & 6,144 & 64.3 & 16 & 1.0 & 24.74 \\
			d=64, $\lambda$=1 & 5.20 & 6,912 & 76.8 & 64 & 1.0 & 25.93 \\
			$\lambda$=0.5 (d=32) & 4.48 & 6,400 & 68.0 & 32 & 0.5 & 24.07 \\
			$\lambda$=2.0 (d=32) & 4.48 & 6,400 & 68.3 & 32 & 2.0 & 25.18 \\
			w/o TARRI & 4.36 & 6,400 & 67.2 & 32 & 1.0 & 22.92 \\
			Prompt Only & 3.67 & 6,144 & 62.1 & - & - & 17.86 \\
			Prompt-guided & 3.52 & 7,280 & 92.4 & - & - & 20.18 \\
			DiffSC & 2.71 & 4,190,000 & 7106 & - & - & 19.65 \\
			DeepJSCC & 7.55 & 21,846 & 454.3 & - & - & 17.71 \\
			BPG & - & 8,120 & 18.4 & - & - & 15.78 \\
			\hline
		\end{tabular}%
	}
	\vspace{-12pt}
\end{table}

Table~\ref{tab:complexity_ablation} summarizes the computational cost and ablation results. The proposed method reduces payload with lower inference time than both baselines. The \emph{Prompt Only} and \emph{w/o TARRI} variants show similar complexity, confirming that the introduced residual transmission and TARRI incur only marginal overhead. The ablation results on residual dimension $d$ and injection strength $\lambda$ indicate that $d=32$ achieves the best performance-complexity trade-off, and $\lambda=1.0$ yields the most robust reconstruction quality, justifying our design choices.

\vspace{-6pt}
\section{Conclusion} 
\label{sec:conclusion}
\vspace{-4pt}

This letter presented a Flow Matching-based generative SemCom framework for robust image transmission under extremely limited payload. The proposed framework jointly transmits a prompt embedding and a compact structural residual velocity vector extracted from the Flow Matching inversion trajectory. Furthermore, a TARRI network was proposed to progressively release the transmitted structural memory along the denoising trajectory, enabling temporally adaptive structural refinement during image reconstruction. Experimental results show that the framework achieves superior reconstruction quality with lower transmission payload and inference latency compared to existing SemCom schemes.

\ifCLASSOPTIONcaptionsoff
\newpage
\fi

\vspace{-4pt}
\bibliographystyle{IEEEtran}
\vspace{-2pt}
\bibliography{refer.bib}

\end{document}